\documentclass[11pt,letterpaper,leqno]{article}

\usepackage[utf8]{inputenc}
\usepackage[english]{babel}
\usepackage{palatino}

\usepackage[margin=.9in]{geometry}
\usepackage{multicol}
\usepackage{ragged2e}
\usepackage{enumitem}
\usepackage{epigraph}
\usepackage{chngcntr}

\usepackage{amsmath}
\usepackage{amssymb}
\usepackage{amsfonts}
\usepackage{amscd}
\usepackage{mathtools}
\usepackage{amsthm}
\usepackage{thmtools}
\usepackage{bm}
\usepackage{bbm}
\usepackage{xfrac}  
\DeclareMathAlphabet{\mathstandardcal}{OMS}{cmsy}{m}{n}

\usepackage[usenames,dvipsnames,svgnames,table]{xcolor}

\definecolor{customblue}{rgb}{0.2235, 0.4157, 0.6941}
\definecolor{customred}{rgb}{0.8000, 0.1451, 0.1608}
\definecolor{ao}{rgb}{0.0, 0.5, 0.0}
\definecolor{afuchs}{rgb}{0.57, 0.36, 0.51}
\definecolor{amber}{rgb}{1, 0.75, 0}
\definecolor{air}{rgb}{0.36, 0.54, 0.66}

\usepackage{graphicx}
\usepackage{epsfig} 
\graphicspath{ {images/} }
\usepackage{float}
\usepackage{placeins}
\usepackage{caption}
\usepackage{subcaption}
\usepackage{multirow}
\usepackage{tabularx}
\usepackage{booktabs}
\usepackage{siunitx} 

\usepackage{tikz}
\usetikzlibrary{calc, intersections, through, backgrounds, shadings, shapes.geometric, arrows.meta, patterns}
\usepackage{pgfplots}
\usepgfplotslibrary{fillbetween}

\usepackage{etoolbox}
\usepackage{apptools}

\usepackage[bottom,multiple]{footmisc}
\usepackage{natbib}
\setcitestyle{authoryear,round,longnamesfirst=false,maxnames=2}

\usepackage{xr-hyper}

\usepackage[hyperfootnotes=false]{hyperref} 
\hypersetup{
    pdfborder=0 0 0,
    colorlinks=true, 
    urlcolor=MidnightBlue, 
    citecolor=MidnightBlue, 
    linkcolor=MidnightBlue,
    filecolor=MidnightBlue
}

\usepackage{nameref}
\usepackage{cleveref}

\theoremstyle{plain}

\newtheorem{prop}{Proposition}

\AtAppendix{\counterwithin{prop}{subsection}}
\AtAppendix{\counterwithin{lem}{subsection}}
\AtAppendix{\counterwithin{cor}{subsection}}
\AtAppendix{\counterwithin{cl}{subsection}}

\newcommand\nc{\newcommand}
\nc\on{\operatorname}

\newcommand{\ba}{\boldsymbol{a}}
\newcommand{\bh}{\boldsymbol{h}}

\newrobustcmd*{\citefirstlastauthor}{\AtNextCite{\DeclareNameAlias{labelname}{given-family}}\citeauthor}

\makeatletter
\def\@footnotecolor{gray!70!black}
\define@key{Hyp}{footnotecolor}{%
 \HyColor@HyperrefColor{#1}\@footnotecolor%
}
\patchcmd{\@footnotemark}{\hyper@linkstart{link}}{\hyper@linkstart{footnote}}{}{}

\renewcommand\@biblabel[1]{}

\makeatother

\hypersetup{footnotecolor=black}

\begin{document}
\title{The Turing Valley: How AI Capabilities Shape Labor Income\thanks{Department of Economics, IESE Business School, Carrer d'Arn\'{u}s i de Gar\'{\i} 3-7, 08034 Barcelona, Spain (eide@iese.edu and etalamas@iese.edu). We thank Pascual Restrepo (the Editor) and two anonymous referees for extremely useful feedback. We have also benefited from the comments of Daron Acemoglu, Ricardo Alonso, Pol Antràs, Jeff Ely, Luis Garicano, Jason Hartline, Johannes Horner, Todd Lensman, Jin Li, Víctor Mart\'{i}nez de Alb\'{e}niz, Gaizka Ormazabal, Crist\'{o}bal Otero, Gerard Padró-i-Miquel, Esteban Rossi-Hansberg, Martin Rotemberg, Raffaella Sadun, Jaume Ventura, Rakesh Vohra and Xavier Vives. We acknowledge the financial support of IESE through the High Impact Initiative-course 2024/2026. We declare we have no relevant or material financial interests that relate to the research described in this paper.}  }

\author{Enrique Ide \& Eduard Talam\`{a}s}
\date{\today}
\maketitle

\begin{abstract}  There is concern that progress toward AI systems with strong capabilities across domains will reduce the importance of human input in production and thus wages. We show that when knowledge is tacit and multidimensional, making AI less jagged can instead raise labor's marginal product. Tacit knowledge makes sequential problem solving efficient because problems cannot be assigned ex ante to the agent best equipped to solve them. When organizations cannot fully integrate human and AI knowledge across dimensions, improving AI where humans initially have an advantage can remove from the referral pool problems that would otherwise consume human time and remain unsolved. By concentrating scarce human time on problems humans can solve,  better screening can raise labor's marginal product even as fewer problems require human input. Our results imply that the human--AI versus AI-only performance gap used to measure human contribution to output need not track the marginal product of labor. \end{abstract}

\newpage 

\section{Introduction}

A central ambition guiding the development of Artificial Intelligence (AI) is to create Artificial General Intelligence (AGI): to replace the ``jagged'' performance profiles of current systems---above human levels in some dimensions and below them in others---with consistently high performance across domains.\footnote{As Demis Hassabis, CEO and co-founder of Google DeepMind, explains \citep{hassabis2025}: ``[AGI] isn't kind of a jagged intelligence where some things it's really good at but other things it's really flawed at. That's what we currently have with today's systems. They're not consistent. So you'd want consistency of intelligence across the board.'' Mathematician Terence Tao expresses a similar idea as follows \citep{tao2026}: ``Right now we're going through a cognitive version of the Copernican revolution,
where we used to think that human intelligence is the center of the universe. And now we're actually seeing that there's very different types of intelligence that are out there with very different strengths and weaknesses.''} This ambition builds on the long-standing objective embodied in the Turing Test \citep{turing1950}. At the same time, there are concerns that progress toward AGI may reduce the importance of human input in production and lower wages \citep{brynjolfsson2022turing,acemoglu2024learning,Restrepo2025Wont}.

Motivated by these concerns, this paper studies how changes in AI capabilities affect labor income when knowledge is tacit and multidimensional. We show that some forms of progress toward more general AI can raise labor's marginal product even as they reduce the set of problems requiring human input. The reason is that labor's marginal product depends not only on the range of problems for which human input is needed, but also on how the organization of production selects the problems on which human time is spent.

We start with the simplest setting that captures the mechanism. All humans share the same capabilities, and all machines run the same AI system. Knowledge and problem difficulty are two-dimensional, and capabilities are jagged: machines are more capable than humans in one dimension and less capable in the other. Production requires solving problems whose difficulty is unknown ex ante, and a problem is solved only if the agent has sufficient knowledge in both dimensions. Each attempt consumes one unit of time, regardless of the outcome. Knowledge is tacit, so an agent learns whether it can solve a problem only by attempting it. Humans and machines can produce independently or in two-layer hierarchies in which machines attempt problems first and refer those they cannot solve to a human. Machine time is abundant relative to human time, so humans capture the surplus generated by the organization through wages.

In this setting, human--AI knowledge hierarchies create value through two channels: screening and knowledge integration. Screening changes the composition of the problems on which scarce human time is spent. By attempting problems first, machines screen out both problems humans could solve and problems they could not. Removing the former makes the referral pool less favorable to humans; removing the latter makes it more favorable. When the latter effect dominates, screening raises the productivity of human time. Knowledge integration instead expands the set of problems the hierarchy can solve. By combining the machine's strength in one dimension with the human's strength in the other, the hierarchy can solve problems that neither agent can solve independently.

We then consider improvements in AI’s weak dimension, which reduce jaggedness by making machines more capable where humans initially have the advantage. The problems newly solvable by AI fall into two groups. One group consists of problems previously referred to and solved by humans. Automating them shrinks the set of problems requiring human input and lowers labor’s marginal product. The remaining problems were previously referred to humans but remained unsolved. Solving them autonomously eliminates unsuccessful referrals and concentrates scarce human time on problems humans can solve, raising labor’s marginal product. When the gains from eliminating unsuccessful referrals outweigh the losses from automating problems previously solved by humans, wages rise even as the set of problems requiring human input contracts.

The extent of knowledge integration determines whether this favorable screening effect can arise. Under perfect integration, the hierarchy can combine human and machine knowledge to solve every newly AI-solvable problem that neither agent can solve independently. An improvement in AI’s weak dimension therefore removes only successful referrals and unambiguously lowers wages. Under imperfect integration, by contrast, some of these problems remain unsolved after referral. Making them autonomously solvable eliminates unsuccessful referrals, leaving scope for less jagged AI to raise wages. Knowledge integration therefore governs whether progress toward more general AI complements labor or merely substitutes for it.\footnote{Knowledge integration appears challenging for current AI systems. As Terence Tao explains \citep{tao2026}: ``These [AI] tools either succeed or they fail. And they've been really bad at creating sort of partial progress or identifying intermediate stages that you should focus on first. [\ldots] We don't have a way of evaluating partial progress the same way we can evaluate a one-shot success or failure of solving a problem.''}

\begin{figure}[!b]
\centering
\begin{tikzpicture}[
    box/.style={
      draw,
      rounded corners=3pt,
      minimum width=3.25cm,
      minimum height=1.35cm,
      align=center,
      inner sep=5pt
    },
    arr/.style={-{Latex[length=2.6mm]},very thick},
    every node/.style={font=\small}
]
  \node[box,fill=gray!10] (U) at (0,0)
    {Previously unsolvable};

  \node[box,fill=customblue!10] (O) at (5.0,0)
    {Solvable only with\\human input};

  \node[box,fill=ao!10] (A) at (10.0,0)
    {AI-solvable};

  \draw[arr,customblue]
    (U.north east)
    to[bend left=18]
    node[
      above=4pt,
      align=center,
      fill=white,
      inner sep=1pt
    ]
    {Augmentation}
    (O.north west);

  \draw[arr,customred]
    (O.north east)
    to[bend left=18]
    node[
      above=4pt,
      align=center,
      fill=white,
      inner sep=1pt
    ]
    {Automation}
    (A.north west);

  \draw[arr,ao]
    (U.south)
    to[bend right=10]
    node[below=4pt,align=center]
    {Better Screening}
    (A.south);
\end{tikzpicture}

\captionsetup{justification=centering}
\caption{Three Problem-Level Transitions Generated by Improvements in AI}
\label{fig:three_transitions}
\end{figure}

The preceding analysis suggests a more general problem-level decomposition of the labor-income effects of AI progress. At any date, each problem is in one of three states: solvable by AI alone, solvable only with human input, or unsolvable by the human--AI organization. As Figure \ref{fig:three_transitions} illustrates, AI improvements can induce three transitions among these states. The first is automation: a problem solvable only with human input becomes solvable by AI alone. By displacing productive human input, automation lowers labor's marginal product. The second is augmentation: a problem unsolvable by the organization becomes solvable only with human input. By expanding the set of problems on which human input is productive, augmentation raises labor's marginal product.

The third transition occurs when a problem that the human--AI organization cannot solve becomes autonomously solvable by AI. It is neither automation nor augmentation: human input was not productive before the improvement and remains unnecessary afterward. Even so, it can raise labor's marginal product because sequential production implies that the second-stage agent faces a selected problem pool, and the improvement raises the value of operating on that pool. When machines move first, the improvement eliminates unsuccessful referrals, making the remaining pool more favorable to humans and raising the productivity of human time. When humans move first, their failures select the problems machines face, so newly AI-solvable problems are overrepresented among referrals. The improvement therefore raises machine productivity more within the organization than in independent production, with the resulting surplus accruing to humans due to the abundance of compute. We refer to the labor-complementing force common to both orderings as \textit{positive screening complementarity}.

Our results also have implications for measuring the contribution of human input to production. A common approach compares the average performance of a human--AI system with that of AI alone on the same distribution of problems \citep[e.g.,][]{cao2024manmachine,vaccaro2024combinations,merali2025scaling}. Under simultaneous problem solving, this performance gap coincides with the marginal product of labor. When tacit knowledge makes sequential problem solving optimal, however, failures at the first stage determine the pool of problems that reaches the second. The marginal product of labor then depends on this selected pool rather than on the unconditional problem distribution used to compute the performance gap.  A narrowing human--AI performance gap therefore need not imply that human input is becoming less valuable in production.

\subsection*{Related Literature}

This paper is most closely related to \citet{AIKE}, who introduce AI into the canonical framework of knowledge hierarchies \citep{garicano2000hierarchies,garicano2004inequality,garicano2006organization}.\footnote{Other important contributions to this literature include \citet{garicanohubbard2,garicano2009specialization,garicano2012growth,bloom2014distinct,fuchs2015optimal,garicano2018earnings,caicedo2019learning,gumpert2022firm,tamayo2024organizational}.} This literature starts from the observation that tacit knowledge makes it difficult to match problems with those who can solve them, giving rise to sequential problem solving as a means of allocating expert time efficiently \citep{garicano2000hierarchies}. The framework is particularly well suited to studying AI because modern systems can acquire and deploy tacit knowledge at scale \citep{brynjolfsson2017can,autor2024applying,brynjolfsson2025generative}. \citet{AIKE} use the framework to study how AI reshapes occupational choice, organizational structure, and the distribution of labor income.

In the canonical one-dimensional knowledge economy, problem-solving capabilities are nested. The less knowledgeable agent attempts problems first and screens out only problems that the more knowledgeable agent could also solve. Referrals are therefore harder than an unconditional problem draw. Hierarchies nevertheless create value because the model assumes that handling a referral takes less time than pursuing a new production opportunity.

In this paper, we deliberately remove this time-saving advantage by assuming that attempting a problem always consumes one unit of time, regardless of an agent’s position in the hierarchy. This allows us to isolate two potential sources of organizational value that arise when knowledge is multidimensional. First, screening can improve the composition of the problems faced by the second-stage agent. Second, knowledge integration can allow the hierarchy to solve problems that neither agent can solve independently. 

Our paper also contributes to the broader literature on automation and technological change \citep[e.g.,][]{zeira1998workers,autor2003skill,acemoglu2018race,acemoglu2019automation,brynjolfsson2022turing,moll2022uneven,acemoglu2024learning,ide2025automation,Restrepo2025Wont} by identifying a novel channel through which improvements in AI capabilities can complement labor. This channel arises when tacit knowledge makes it difficult to match problems to those who can solve them, giving rise to sequential problem solving. In this setting, AI progress can raise the marginal product of labor by reducing the resources devoted to problems that would otherwise remain unsolved.

This channel is distinct from the forces emphasized in task-based models of technological change. \citet{acemoglu2018race} contrast automation---which transfers existing tasks from labor to capital and generates both displacement and productivity effects---with the creation of new labor-intensive tasks. \citet{acemogluKongRestrepo2024Tasks} develop a broader taxonomy that also includes labor- and capital-augmenting technological change. The screening channel we identify falls outside this taxonomy: it does not transfer an existing task from labor to AI, exogenously increase a factor’s productivity in tasks it already performs, or create new labor-intensive tasks. Instead, it raises labor’s marginal product by improving how organizations allocate scarce human and machine time across problems. To our knowledge, this screening-based channel has no parallel in the existing automation literature.

Finally, we contribute to the growing literature on the empirical evaluation of human--AI collaboration \citep[e.g.,][]{noy2023experimental,cao2024manmachine,vaccaro2024combinations,brynjolfsson2025generative,dell2026cybernetic,dellacqua2026jagged}. Among these papers, the one most closely related to our analysis is \citet{merali2025scaling}. The study assigns more than five hundred professionals to complete management, consulting, and data-analysis tasks with one of thirteen AI models, allowing economically meaningful productivity outcomes to be compared across models with different capabilities. It finds that AI-only performance improves with model scale, while human--AI performance remains comparatively flat, narrowing the human--AI performance gap. Our framework shows why this narrowing need not imply a decline in labor's marginal product.

\vspace{3mm}
\noindent\textit{Roadmap.}---Section~\ref{sec:hierarchies} presents the simplest setting that captures our main mechanism and characterizes the value of human--AI hierarchies. Section~\ref{sec:turing_valley} derives the effects of AI improvements on labor income under varying degrees of knowledge integration. Section~\ref{sec:decomposition_measurement} develops a general problem-level decomposition. Section~\ref{sec:measurement} examines the implications for measurement. Section~\ref{sec:final} concludes.

\section{The Value of Hierarchies in a Multidimensional Knowledge Economy} \label{sec:hierarchies}

This section shows how human--AI hierarchies create value when knowledge is tacit and multidimensional. Tacit knowledge prevents production problems from being assigned ex ante to the agent best equipped to solve them, giving the hierarchy a screening role. Multidimensional knowledge allows humans and machines to be more knowledgeable in different dimensions, so each can solve problems the other cannot. As a result, screening can improve the composition of the problems referred through the hierarchy. In addition, integrating human and AI knowledge can allow the organization to solve problems neither can solve independently. 

These two sources of gains---screening and integration---are absent in the canonical one-dimensional knowledge economy \citep[e.g.,][]{garicano2004inequality,garicano2006organization,AIKE}, where agents' problem-solving capabilities are nested. 

\subsection{The Screening Value of Hierarchies} \label{sec:screening}

There is a unit mass of identical humans, each endowed with one unit of time and exogenous knowledge $\boldsymbol{h}=(h_1,h_2)\in(0,1)^2$. There are also $\mu>0$ units of computing power---or ``compute''---normalized so that one unit of compute is equivalent to one unit of time. An exogenous AI algorithm equips every unit of compute with the same knowledge $\boldsymbol{a}=(a_1,a_2) \in [0,1]^2$. For simplicity, we refer to one unit of compute as a ``machine,'' with each machine having one ``unit of time.''

Motivated by the jagged nature of current AI systems \citep{Moravec1988MindChildren,brynjolfsson2022turing}, we focus on the case in which AI outperforms humans in one dimension but underperforms them in the other:\begin{equation*}
a_1>h_1
\qquad\text{and}\qquad
a_2<h_2.
\end{equation*}Thus, unlike in the canonical one-dimensional knowledge economy, agents’ knowledge profiles cannot be ranked.

Production requires pursuing production opportunities. Each opportunity is identical ex ante and, once pursued, presents a problem with difficulty $\boldsymbol{x}\equiv(x_1,x_2)$, drawn from a distribution with full support on $[0,1]^2$ and cumulative distribution function $F$. Pursuing an opportunity requires one unit of time, irrespective of whether the problem is ultimately solved. An agent with knowledge $\boldsymbol{z}\equiv(z_1,z_2)$ solves the problem if $z_1\geq x_1$ and $z_2\geq x_2$. A solution produces one unit of output; an unsolvable problem produces none. We normalize the output price to one, allowing us to measure income in units of output. 

By relabeling machines as a second type of human specialist with knowledge $\boldsymbol{a}$, or humans with a second type of AI with knowledge $\boldsymbol{h}$, the same model could describe an economy with two types of human specialists or AI models, respectively. We adopt the human--AI interpretation because capability differences between humans and AI are arguably more jagged across dimensions than those among human specialists or among AI models.\footnote{Describing the unconventional playing style of DeepMind’s AlphaZero, Demis Hassabis observed that it ``doesn’t play like a human, and it doesn’t play like a program. It plays in a third, almost alien, way''  \citep{knight2017alpha}.} The human--AI setting is also particularly suited to our central question---how improvements in one agent's capabilities affect the income of the other---because AI capabilities are improving rapidly and unevenly \citep{brynjolfsson2022turing}.

When humans and machines operate independently, each uses its unit of time to pursue a production opportunity. A human solves the associated problem with probability $F(\boldsymbol{h})$, whereas a machine does so with probability $F(\boldsymbol{a})$. Their expected outputs---and therefore their respective autarky incomes---are $F(\boldsymbol{h})$ and $F(\boldsymbol{a})$. These success probabilities can be characterized using the partition of the problem space shown in Figure \ref{fig:1}. Region $B$ contains problems that both agents can solve; region $H$ those that only the human can solve; region $A$ those that only the machine can solve; and region $N$ those that neither can solve. Thus,
\[ F(\boldsymbol{h})=\Pr(\boldsymbol{x}\in B\cup H) \qquad\text{and}\qquad F(\boldsymbol{a})=\Pr(\boldsymbol{x}\in B\cup A) \]

\begin{figure}[t!]
\centering
\pgfmathsetmacro{\hone}{0.35}   
\pgfmathsetmacro{\htwo}{0.75}   
\pgfmathsetmacro{\mone}{0.75}   
\pgfmathsetmacro{\mtwo}{0.35}   
 \begin{tikzpicture}[scale=1.05]
        \pgfmathsetlengthmacro\MajorTickLength{
       \pgfkeysvalueof{/pgfplots/major tick length} *1.5
     }
\begin{axis}[ytick={\mtwo, \htwo, 1},
         yticklabels={$a_2$, $h_2$, 1},
         xtick={0, \hone, \mone, 1},
         xticklabels={0, $h_1$ ,$a_1$, 1},
         xmin=0,
     xmax=1,
     ymin=0,
     ymax=1,
     axis line style=thick,
     yticklabel style={
    fill=white,
    },
    ylabel={Dimension 2},
    xticklabel style={
    fill=white
    },
    xlabel={Dimension 1},
    ylabel near ticks,
    xlabel near ticks,
    ylabel style={rotate=-0},
    major tick length=\MajorTickLength,
      every tick/.style={
        black,
        thick,
      },]
      
        \fill[pattern=north west lines, pattern color=gray!25] (axis cs:0,0) -- (axis cs:0,\htwo) -- (axis cs:\hone,\htwo) -- (axis cs:\hone,\mtwo) -- (axis cs:\mone,\mtwo) -- (axis cs:\mone,0) -- cycle;

\addplot[name path=ceiling,domain={0:1},opacity=0] {1};        
\addplot[name path=floor,domain={0:1},opacity=0] {0};  
       \draw[thick, black, dotted] (axis cs:0,\mtwo) -- (axis cs:\mone,\mtwo) -- (axis cs:\mone,0);
        \draw[thick, black,  dotted] (axis cs:0,\htwo) -- (axis cs:\hone,\htwo) -- (axis cs:\hone,0);

         \draw[thick, black] (axis cs:\hone,\mtwo) -- (axis cs:\mone,\mtwo);
         \draw[thick, black] (axis cs:0,\htwo) -- (axis cs:\hone,\htwo);
          \draw[thick, black] (axis cs:\hone,\htwo) -- (axis cs:\hone,\mtwo);
           \draw[thick, black] (axis cs:\mone,\mtwo) -- (axis cs:\mone,0);

        \node[font=\Large] at (axis cs:0.15,0.55) {$H$};
        \node[font=\Large] at (axis cs:0.15,0.15) {$B$};
        \node[font=\Large] at (axis cs:0.55,0.15) {$A$};
                \node[font=\Large] at (axis cs:0.7,0.7) {$N$};
             
\end{axis}
\end{tikzpicture} 

\captionsetup{justification=centering}
\caption{ Human and Machine Problem-Solving Capabilities \\ \justifying 
\footnotesize \noindent \textit{Notes}. The figure partitions the problem space according to whether a human with knowledge $\boldsymbol{h}$ or a machine with knowledge $\boldsymbol{a}$ can solve each problem. Region $B$ contains problems that both can solve; regions $H$ and $A$ contain problems that only the human and only the machine can solve, respectively; and region $N$ contains problems that neither can solve.}\label{fig:1}
\end{figure}

We next ask whether humans and machines can create additional value by organizing production jointly. The first organizational choice is whether they attempt problems sequentially or simultaneously. Under sequential problem solving, one agent attempts each problem first and refers failures to the other; under simultaneous problem solving, both agents attempt the same problem from the outset. Appendix \ref{app:importance} shows that sequential problem solving yields strictly higher output under fairly general conditions that hold in all the environments studied here. The gain rests on the fact that knowledge is tacit: an agent learns whether it can solve a problem only by attempting it, so problems cannot be assigned ex ante to the agent best equipped to solve them. Sequential problem-solving therefore avoids wasting the second agent’s time and knowledge on problems the first can solve. We accordingly restrict attention in the main text to sequential problem solving.

Four organizational features, however, remain to be specified: which agent attempts problems first; how much time the second agent spends handling each referral; whether human time or compute limits the number of hierarchies that can form; and whether the organization can integrate human and machine knowledge. The first three features also arise in one-dimensional knowledge hierarchies with AI \citep[e.g.,][]{AIKE,GKW}; the fourth is specific to a multidimensional knowledge economy.

For expositional clarity, we focus on the following benchmark. Machines tackle problems first and refer failures to humans. Each referral consumes one full unit of human time. Compute is abundant relative to human time, so human time is the bottleneck limiting the number of hierarchies that can form. Finally, partial solutions cannot be integrated across dimensions, so a human who receives a referral can solve it only if she can solve it independently. The first three assumptions are not essential to our main insights.\footnote{We discuss the role of referral time below and the implications of compute scarcity in footnote~\ref{foot:abundance}. Section~\ref{sec:decomposition_measurement} discusses the case in which machines tackle problems second, and Appendix~\ref{app:machines_second} analyzes it formally.} 
Section~\ref{sec:integration} relaxes the final assumption by allowing human and machine knowledge to be integrated across dimensions.

Because each machine generates a referral with probability $1-F(\boldsymbol{a})$, and each referral consumes one unit of human time, fully utilizing a human's unit of time requires matching her with $n(\boldsymbol{a})$ machines, where:\footnote{Following \cite{garicano2000hierarchies} and subsequent work, we treat span of control as continuous and abstract from integer constraints. Formally, the one-human description is shorthand for a positive mass of identical humans matched with $n(\boldsymbol a)$ times as many machines, with referrals pooled across humans. An exact law of large numbers then makes the aggregate mass of referrals equal to its expectation. This convention abstracts from queueing and idle time caused by stochastic referral arrivals in finite teams. Since the mass of humans is one, compute is abundant when $\mu>n(\boldsymbol a)$.}
\[
n(\boldsymbol{a})
\equiv
\frac{1}{1-F(\boldsymbol{a})}.
\]The human therefore handles one problem, as in autarky; the hierarchy changes which problems she handles, not how many. This contrasts with the canonical one-dimensional knowledge economy, where handling a referral takes only $c<1$ units of time, allowing agents on the top layer to handle $1/c>1$ referrals per unit of time.\footnote{One interpretation of this assumption is that production involves two time-consuming activities: pursuing an opportunity---for example, drafting a memo or preparing slides---and solving the underlying problem---for example, deciding what the memo should say or how the slides should be organized. By focusing exclusively on problem solving---while directing the bottom layer to execute those solutions---the top-layer agent saves time and can engage with $1/c>1$ problems rather than pursue and solve a single opportunity from start to finish.} There, hierarchies create value by increasing the number of problems those agents try to solve. By setting $c=1$, we shut down this traditional value of hierarchies and isolate the novel organizational gains uncovered in this paper. 

The organization succeeds on an opportunity if the machine solves the associated problem or, following a referral, the human solves it independently. Its solvable set is therefore $B\cup H\cup A$ in Figure \ref{fig:1}, and its probability of success on each opportunity is:\[ \Phi^N(\boldsymbol{h},\boldsymbol{a}) \equiv \Pr(\boldsymbol{x}\in B\cup H\cup A) =F(\boldsymbol{h})+F(\boldsymbol{a})-F(\boldsymbol{h}\wedge\boldsymbol{a}) \]
where $\boldsymbol{h}\wedge\boldsymbol{a}\equiv(h_1,a_2)$ is the coordinate-wise minimum of the human and machine knowledge profiles, so $F(\boldsymbol{h}\wedge\boldsymbol{a})=\Pr(\boldsymbol{x}\in B)$. Since a hierarchy combines one human with $n(\boldsymbol{a})$ machines, each of which pursues one production opportunity, its total expected output is $n(\boldsymbol{a})\Phi^N(\boldsymbol{h},\boldsymbol{a})$. 

To characterize how this organizational output is divided between humans and machines, let $r$ denote the equilibrium rental rate of a machine, and $w^N$ denote the wage offered by a no-integration hierarchy. Because compute is abundant, some machines operate independently in equilibrium, pinning their rental rate to their autarky return:
\[ r=F(\boldsymbol{a}) \]A hierarchy must therefore pay $n(\boldsymbol{a})r$ for the machines it rents. The remaining output accrues to the human as wages, so:\[
w^N
=
n(\boldsymbol{a})
\left[
\Phi^N(\boldsymbol{h},\boldsymbol{a})-r
\right]  =
\frac{
\Pr(\boldsymbol{x}\in H)
}{
1-\Pr(\boldsymbol{x}\in A\cup B)
} = \Pr(\text{Human succeeds} \, | \, \text{AI fails})
\]The hierarchy wage $w^N$ therefore equals the probability that the human solves a problem conditional on the machine being unable to do so.

A human operating in autarky earns $F(\boldsymbol{h})=\Pr(\boldsymbol{x}\in B\cup H)=\Pr(\text{Human succeeds})$.  The hierarchy therefore creates value---and that value accrues to the human\footnote{Whether human time or compute is scarce determines who captures the hierarchy's surplus. When compute is abundant relative to human time, some machines operate independently and the surplus accrues to the human; when compute is scarce, some humans operate independently and the surplus accrues to machine owners. We view compute abundance as the natural benchmark because it captures settings in which human time is the bottleneck limiting the number of hierarchies that can form. Under this benchmark, the organizational-value effects studied below coincide with wage effects; with scarce compute, the analysis instead concerns organizational value rather than labor income. \label{foot:abundance}}---whenever $w^N>F(\boldsymbol{h})$, or equivalently whenever:\begin{equation} \label{eq:key} \Pr(\text{Human succeeds} \, | \, \text{AI fails}) > \Pr(\text{Human succeeds}).
\end{equation}Thus, the hierarchy is valuable when the problems referred by machines are more favorable to the human than an unconditional problem draw. If the inequality is reversed, the human instead pursues a new production opportunity and earns her autarky income $F(\boldsymbol{h})$.

The contrast with the post-AI one-dimensional knowledge economy is immediate. There, agents’ knowledge profiles can be ranked, so the machine’s problem-solving capabilities are nested within those of the more knowledgeable human who handles its referrals. Every problem the machine solves is therefore one the human could also solve. The machine thus screens out only problems the human would have solved, making the referral pool less favorable to her than an unconditional problem draw. Consequently, hierarchies can arise in equilibrium only if handling a referral takes less time than pursuing a new production opportunity (i.e., $c<1$).

Multidimensional knowledge breaks this logic because human and machine capabilities need not be nested. As Figure \ref{fig:1} illustrates, the machine screens out problems in both $B$, which the human can solve, and $A$, which she cannot. Screening out problems in $B$ makes the referral pool less favorable to the human, whereas screening out problems in $A$ makes it more favorable. When the favorable screening effect from region $A$ dominates the adverse effect from region $B$---that is, when condition (\ref{eq:key}) holds---the hierarchy creates value even though humans tackle no more problems than in autarky. We call this gain the \textit{positive screening value} of multidimensional knowledge hierarchies.

This screening value is inherently tied to tacit knowledge, which makes sequential problem solving optimal. The machine’s attempts then determine the pool of problems on which human time is spent, and the composition of that pool governs the productivity of human time.

\subsection{The Integration Value of Organizations} \label{sec:integration}

Multidimensional knowledge gives rise to a second source of organizational value: knowledge integration. By combining the machine's strength in one dimension with the human's strength in the other, the organization can solve problems that neither agent can solve independently. For expositional clarity, we first study the polar case of perfect integration and return to imperfect integration at the end of Section~\ref{subsec:integration}.

Formally, let $\boldsymbol{h}\vee\boldsymbol{a}$ denote the coordinate-wise maximum of the human and machine knowledge profiles. Because $a_1>h_1$ and $h_2>a_2$, we have $\boldsymbol{h}\vee\boldsymbol{a}=(a_1,h_2)$. Under perfect integration, the organization solves a problem if and only if $\boldsymbol{x}\leq\boldsymbol{h}\vee\boldsymbol{a}$, where the inequality is understood coordinate-wise. Its probability of success on each production opportunity is therefore:\[ \Phi^I(\boldsymbol{h},\boldsymbol{a}) \equiv F(\boldsymbol{h}\vee\boldsymbol{a}) \]

\begin{figure}[t!]
\centering
\pgfmathsetmacro{\hone}{0.35}   
\pgfmathsetmacro{\htwo}{0.75}   
\pgfmathsetmacro{\mone}{0.75}   
\pgfmathsetmacro{\mtwo}{0.35}   
 \begin{tikzpicture}[scale=1.05]
        \pgfmathsetlengthmacro\MajorTickLength{
       \pgfkeysvalueof{/pgfplots/major tick length} *1.5
     }
\begin{axis}[ytick={\mtwo, \htwo, 1},
         yticklabels={$a_2$, $h_2$, 1},
         xtick={0, \hone, \mone, 1},
         xticklabels={0, $h_1$ ,$a_1$, 1},
         xmin=0,
     xmax=1,
     ymin=0,
     ymax=1,
     axis line style=thick,
     yticklabel style={
    fill=white,
    },
    ylabel={Dimension 2},
    xticklabel style={
    fill=white
    },
    xlabel={Dimension 1},
    ylabel near ticks,
    xlabel near ticks,
    ylabel style={rotate=-0},
    major tick length=\MajorTickLength,
      every tick/.style={
        black,
        thick,
      },]
      
        \fill[pattern=north west lines, pattern color=gray!25] (axis cs:0,0) -- (axis cs:0,\htwo) -- (axis cs:\hone,\htwo) -- (axis cs:\hone,\mtwo) -- (axis cs:\mone,\mtwo) -- (axis cs:\mone,0) -- cycle;
         \fill[pattern=north east lines, pattern color=ao!50] (axis cs:\hone,\mtwo) rectangle (axis cs:\mone,\htwo);

\addplot[name path=ceiling,domain={0:1},opacity=0] {1};        
\addplot[name path=floor,domain={0:1},opacity=0] {0};

       \draw[thick, black, dotted] (axis cs:0,\mtwo) -- (axis cs:\mone,\mtwo) -- (axis cs:\mone,0);
        \draw[thick, black,  dotted] (axis cs:0,\htwo) -- (axis cs:\hone,\htwo) -- (axis cs:\hone,0);

         \draw[thick, black] (axis cs:0,\htwo) -- (axis cs:\mone,\htwo);
           \draw[thick, black] (axis cs:\mone,\htwo) -- (axis cs:\mone,0);

        \node[font=\Large] at (axis cs:0.15,0.55) {$H$};
        \node[font=\Large] at (axis cs:0.15,0.15) {$B$};
        \node[font=\Large] at (axis cs:0.55,0.15) {$A$};
         \node[font=\Large] at (axis cs:0.8,0.85) {$N \setminus HA$};
         \node[font=\Large] at (axis cs:0.55,0.55) {$HA$};

\end{axis}
\end{tikzpicture} 

\captionsetup{justification=centering}
\caption{Problem-Solving Capabilities under Perfect Integration \\ \justifying 
\footnotesize \noindent \textit{Notes}. The figure partitions the problem space according to whether a human with knowledge $\boldsymbol{h}$, a machine with knowledge $\boldsymbol{a}$, or the human--machine organization can solve each problem. Region $B$ contains problems that both the human and the machine can solve; regions $H$ and $A$ contain problems that only the human and only the machine can solve, respectively; and region $N$ contains problems that neither can solve independently. Under perfect integration, the organization can solve the subset $HA\subset N$ by integrating human and machine knowledge across dimensions, while the problems in $N\setminus HA$ remain unsolvable.} \label{fig:2}
\end{figure}

Figure \ref{fig:2} gives the geometric interpretation. Relative to the no-integration benchmark, perfect integration expands the organization's solvable set from $(B\cup H\cup A)$ to $(B\cup H\cup A\cup HA)$. Region $HA$ contains problems that neither agent can solve independently, but for opposite reasons: the human lacks sufficient knowledge in dimension 1, while the machine lacks sufficient knowledge in dimension 2. By integrating the machine's knowledge in dimension 1 with the human's knowledge in dimension 2, the organization can solve these problems. 

Perfect integration changes what the organization can accomplish after a problem is referred, but not which problems are referred. The referral rate therefore remains the same, so each human continues to be matched with $n(\boldsymbol a)$ machines and to receive one referral. Compute abundance also continues to pin the machine rental rate at $r=F(\boldsymbol a)$. Letting $w^I$ denote the wage a perfect-integration hierarchy can offer, we therefore have that: \[
w^I
=
n(\boldsymbol{a})
\left[
\Phi^I(\boldsymbol{h},\boldsymbol{a})-r
\right]
=
\frac{
\Pr(\boldsymbol{x}\in H\cup HA)
}{
1-\Pr(\boldsymbol{x}\in A\cup B)
} = \Pr(\text{Organization succeeds} \, | \, \text{AI fails})
\]As in the no-integration case, the hierarchy wage equals the probability of success conditional on referral. The key difference is that perfect integration expands the set of successful referrals from $H$ to $H\cup HA$.

The hierarchy is strictly optimal whenever $w^I>F(\boldsymbol{h})$. The value of a perfect-integration hierarchy therefore decomposes into:
\[
w^I-F(\boldsymbol{h})
=
\underbrace{w^N-F(\boldsymbol{h})}_{\text{screening value}}
+
\underbrace{w^I-w^N}_{\text{integration value}}.
\]
The first term on the right-hand side is the screening value characterized in Section \ref{sec:screening}. The second captures the additional value created by integration:
\[
w^I-w^N
=
\frac{
\Phi^I(\boldsymbol{h},\boldsymbol{a})
-
\Phi^N(\boldsymbol{h},\boldsymbol{a})
}{
1-F(\boldsymbol{a})
}
=
\frac{
\Pr(\boldsymbol{x}\in HA)
}{
1-\Pr(\boldsymbol{x}\in A\cup B)
}
>0.
\]
The integration value is therefore positive and can offset a negative screening value, potentially making the perfect-integration hierarchy valuable even when the no-integration hierarchy is not.

Unlike screening, knowledge integration can create value without tacit knowledge or sequential referral, including in horizontal teams that combine specialists’ complementary knowledge or skills \citep[e.g.,][]{demsetz1988theory,freund2025superstar,dell2026cybernetic}.\footnote{Indeed, Appendix \ref{app:importance} shows that integration can make human--AI teams efficient in our setting even when the human and AI are constrained to attempt problems simultaneously---a restriction that eliminates any screening benefit.}  In our setting, integration is layered onto the screening generated by sequential problem solving. For a fixed capability profile, the screening and integration values are distinct. As Section~\ref{sec:turing_valley} shows, however, they interact as AI capabilities improve: integration determines whether the problems AI learns to solve autonomously were previously successful or unsuccessful referrals.

Despite their different organizational origins, both gains require multidimensional, non-nested knowledge. In the canonical one-dimensional knowledge economy, nested problem-solving capabilities imply that screening cannot improve the referral pool and that integrating different agents' knowledge cannot expand the organization’s solvable set. Multidimensional knowledge instead allows hierarchies to create value through favorable screening and integration.

\section{How Changes in AI Capabilities Affect Labor Income} \label{sec:turing_valley}

Having identified how human--AI hierarchies create value, we now study how improvements in AI capabilities affect labor income. We say that an AI improvement is  \emph{labor-complementing} if it raises the equilibrium wage and \emph{labor-substituting} if it lowers it.  

A natural intuition is that progress in AI’s strong dimension further differentiates AI from human knowledge and is therefore labor-complementing, whereas progress in its weak dimension brings AI closer to human knowledge and is therefore labor-substituting. Indeed, our model confirms this intuition when the human and AI are constrained to attempt each problem simultaneously (see Appendix \ref{app:importance}). As we show in this section, however, under sequential problem solving the intuition survives only with perfect integration: without integration, progress in AI’s weak dimension can instead be labor-complementing. Thus, when tacit knowledge makes sequential problem solving optimal, knowledge integration is key to the labor-income consequences of AI progress.\footnote{AI progress may also take the form of improved knowledge integration rather than an increase in $a_1$ or $a_2$. Holding capabilities fixed, better integration leaves autonomous machine performance and referrals unchanged, but converts some failed referrals into successful ones, thereby raising the wage under compute abundance. Since this effect is immediate, we focus on changes in the machine's autonomous capabilities.}

Throughout this section, we maintain the assumptions of Section \ref{sec:hierarchies} and focus on improvements under which capabilities remain jagged, compute abundant, and the hierarchy remains strictly optimal.\footnote{Because these conditions are defined by strict inequalities, they continue to hold after sufficiently small improvements whenever they hold initially.} The hierarchy wage therefore coincides with the equilibrium wage before and after each improvement. To make the changing coordinate explicit, we also write $w^N(a_1,a_2)$ and $w^I(a_1,a_2)$.

\subsection{No Integration} \label{subsec:no_integration}

We begin with the no-integration regime. As shown in Section \ref{sec:screening}, the wage is given by: \[
w^N(a_1,a_2) =
\frac{
\Pr(\boldsymbol{x}\in H)
}{
1-\Pr(\boldsymbol{x}\in A\cup B)
} = \Pr(\boldsymbol{x}\in H \, | \, \text{AI fails})
\]

\begin{figure}[t!]
\centering
\begin{subfigure}{.455\linewidth}
\centering
\pgfmathsetmacro{\hone}{0.35}
\pgfmathsetmacro{\htwo}{0.75}
\pgfmathsetmacro{\mone}{0.75}
\pgfmathsetmacro{\mtwo}{0.35}
\pgfmathsetmacro{\monep}{0.90}
\begin{tikzpicture}[scale=0.875]
\pgfmathsetlengthmacro\MajorTickLength{
  \pgfkeysvalueof{/pgfplots/major tick length} *1.5
}
\begin{axis}[
    ytick={\mtwo,\htwo,1},
    yticklabels={$a_2$,$h_2$,1},
    xtick={0,\hone,\mone,\monep,1},
    xticklabels={0,$h_1 \color{white}' \color{black}$,$a_1\color{white}' \color{black}$,$a_1'$,1},
    xmin=0,xmax=1,ymin=0,ymax=1,
    axis line style=thick,
    yticklabel style={fill=white},
    ylabel={Dimension 2},
    xticklabel style={fill=white},
    xlabel={Dimension 1},
    ylabel near ticks,
    xlabel near ticks,
    ylabel style={rotate=-0},
    major tick length=\MajorTickLength,
    every tick/.style={black,thick},
    clip=false
]

\fill[pattern=north west lines,pattern color=gray!25]
    (axis cs:0,0)--(axis cs:0,\htwo)--(axis cs:\hone,\htwo)--
    (axis cs:\hone,\mtwo)--(axis cs:\mone,\mtwo)--
    (axis cs:\mone,0)--cycle;

\fill[pattern=crosshatch,pattern color=customblue!50]
    (axis cs:\mone,0) rectangle (axis cs:\monep,\mtwo);
\draw[thick,black,dashed]
    (axis cs:\mone,\mtwo)--(axis cs:\monep,\mtwo)--(axis cs:\monep,0);
\draw[thick,black,dashed,->]
    (axis cs:\mone,\mtwo+0.045)--(axis cs:\monep,\mtwo+0.045);
    
\addplot[name path=ceiling,domain={0:1},opacity=0] {1};
\addplot[name path=floor,domain={0:1},opacity=0] {0};

\draw[thick,black,dotted]
    (axis cs:0,\mtwo)--(axis cs:\mone,\mtwo)--(axis cs:\mone,0);
\draw[thick,black,dotted]
    (axis cs:0,\htwo)--(axis cs:\hone,\htwo)--(axis cs:\hone,0);

\draw[thick,black] (axis cs:\hone,\mtwo)--(axis cs:\mone,\mtwo);
\draw[thick,black] (axis cs:0,\htwo)--(axis cs:\hone,\htwo);
\draw[thick,black] (axis cs:\hone,\htwo)--(axis cs:\hone,\mtwo);
\draw[thick,black] (axis cs:\mone,\mtwo)--(axis cs:\mone,0);

\node[font=\Large] at (axis cs:0.15,0.55) {$H$};
\node[font=\Large] at (axis cs:0.15,0.15) {$B$};
\node[font=\Large] at (axis cs:0.55,0.15) {$A$};
\node[font=\Large] at (axis cs:0.70,0.70) {$N$};

\node[font=\large,fill=white,inner sep=1.5pt]
    at (axis cs:{(\mone+\monep)/2},{\mtwo/2}) {$\Delta A$};
\end{axis}
\end{tikzpicture}
\caption*{$\quad \ \ $ (a) Improvement in AI’s Strong Dimension}
\end{subfigure}
\begin{subfigure}{.5\linewidth}
\centering
\pgfmathsetmacro{\hone}{0.35}
\pgfmathsetmacro{\htwo}{0.75}
\pgfmathsetmacro{\mone}{0.75}
\pgfmathsetmacro{\mtwo}{0.35}
\pgfmathsetmacro{\mtwop}{0.52}
\begin{tikzpicture}[scale=0.875]
\pgfmathsetlengthmacro\MajorTickLength{
  \pgfkeysvalueof{/pgfplots/major tick length} *1.5
}
\begin{axis}[
    ytick={\mtwo,\mtwop,\htwo,1},
    yticklabels={$a_2$,$a_2'$,$h_2$,1},
    xtick={0,\hone,\mone,1},
    xticklabels={0,$h_1 \color{white}' \color{black}$,$a_1\color{white}' \color{black}$,1},
    xmin=0,xmax=1,ymin=0,ymax=1,
    axis line style=thick,
    yticklabel style={fill=white},
    ylabel={Dimension 2},
    xticklabel style={fill=white},
    xlabel={Dimension 1},
    ylabel near ticks,
    xlabel near ticks,
    ylabel style={rotate=-0},
    major tick length=\MajorTickLength,
    every tick/.style={black,thick},
    clip=false
]

\fill[pattern=north west lines,pattern color=gray!25]
    (axis cs:0,0)--(axis cs:0,\htwo)--(axis cs:\hone,\htwo)--
    (axis cs:\hone,\mtwo)--(axis cs:\mone,\mtwo)--
    (axis cs:\mone,0)--cycle;

\fill[pattern=dots,pattern color=customred!50]
    (axis cs:0,\mtwo) rectangle (axis cs:\hone,\mtwop);
\fill[pattern=crosshatch,pattern color=customblue!50]
    (axis cs:\hone,\mtwo) rectangle (axis cs:\mone,\mtwop);
    
\addplot[name path=ceiling,domain={0:1},opacity=0] {1};
\addplot[name path=floor,domain={0:1},opacity=0] {0};

\draw[thick,black,dotted]
    (axis cs:0,\mtwo)--(axis cs:\mone,\mtwo)--(axis cs:\mone,0);
\draw[thick,black,dotted]
    (axis cs:0,\htwo)--(axis cs:\hone,\htwo)--(axis cs:\hone,0);

\draw[thick,black] (axis cs:\hone,\mtwo)--(axis cs:\mone,\mtwo);
\draw[thick,black] (axis cs:0,\htwo)--(axis cs:\hone,\htwo);
\draw[thick,black] (axis cs:\hone,\htwo)--(axis cs:\hone,\mtwo);
\draw[thick,black] (axis cs:\mone,\mtwo)--(axis cs:\mone,0);

\draw[thick,black,dashed]
    (axis cs:0,\mtwop)--(axis cs:\hone,\mtwop);
\draw[thick,black,dashed]
    (axis cs:\hone,\mtwop)--(axis cs:\mone,\mtwop)--(axis cs:\mone,0);
\draw[thick,black,dashed,->]
    (axis cs:\mone+0.045,\mtwo)--(axis cs:\mone+0.045,\mtwop);

\node[font=\large,fill=white, inner sep=1.5pt]
    at (axis cs:{\hone/2},{(\mtwo+\mtwop)/2}) {$\Delta B$};
\node[font=\large,fill=white,inner sep=1.5pt]
    at (axis cs:{(\hone+\mone)/2},{(\mtwo+\mtwop)/2}) {$\Delta A$};
   
\node[font=\Large] at (axis cs:0.15,0.61) {$H$};
\node[font=\Large] at (axis cs:0.15,0.15) {$B$};
\node[font=\Large] at (axis cs:0.55,0.15) {$A$};
\node[font=\Large] at (axis cs:0.70,0.70) {$N$}; 
    
\end{axis}
\end{tikzpicture}
\caption*{$\quad \ \ $ (b) Improvement in AI’s Weak Dimension}
\end{subfigure}
\captionsetup{justification=centering}
\caption{The Wage Effects of AI Improvements under No Integration \\ \justifying 
\footnotesize \noindent \textit{Notes}. The figure shows how improvements in AI capabilities change the problem-space partition under no integration. In panel (a), an increase in $a_1$ to $a_1'>a_1$ moves the problems in $\Delta A$, defined by $a_1<x_1\leq a_1'$ and $x_2\leq a_2$, from $N$ to $A$. In panel (b), an increase in $a_2$ to $a_2'>a_2$ moves the problems in $\Delta B$, defined by $x_1\leq h_1$ and $a_2<x_2\leq a_2'$, from $H$ to $B$, and the problems in $\Delta A$, defined by $h_1<x_1\leq a_1$ and $a_2<x_2\leq a_2'$, from $N$ to $A$. All other regions remain unchanged.}
\label{fig:3}
\end{figure}

Consider first an improvement in AI's strong dimension: $a_1$ increases to $a'_1>a_1$, while $a_2$ remains unchanged. As panel (a) of Figure~\ref{fig:3} illustrates, the problems in $\Delta A$, defined by $a_1<x_1\leq a'_1$ and $x_2\leq a_2$, move from $N$ to $A$. Before the improvement, the machine referred these problems, but the human could not solve them; afterward, the machine solves them independently.  The improvement therefore removes from the referral pool only problems that a human would have failed to solve, raising the success rate among the remaining referrals and hence her wage. Thus, improvements in AI's strong dimension are labor-complementing.  

By contrast, consider an improvement in AI's weak dimension: $a_2$ increases to $a'_2\in(a_2,h_2)$, while $a_1$ remains unchanged. As panel (b) of Figure~\ref{fig:3} illustrates, this improvement has two opposing effects on the referral pool. The problems in $\Delta B$, defined by $x_1\leq h_1$ and $a_2<x_2\leq a'_2$, move from $H$ to $B$. Before the improvement, they were referred to and solved by the human; afterward, the machine solves them independently. Their removal makes the referral pool less favorable to the human.

The problems in $\Delta A$, defined by $h_1<x_1\leq a_1$ and $a_2<x_2\leq a'_2$, instead move from $N$ to $A$. Before the improvement, they were referred to the human but remained unsolved; afterward, the machine solves them independently. Their removal makes the referral pool more favorable to the human. Thus, although an improvement in AI’s weak dimension unambiguously shrinks the set of problems solvable only with human input, its wage effect is ambiguous: the wage rises when the favorable screening effect from $\Delta A$ outweighs the adverse screening effect from $\Delta B$.

We formalize these results in the following proposition:

\begin{samepage}
\begin{prop}[AI Improvements - No Integration] \label{prop:no-integration}
Suppose $a_1'>a_1>h_1$ and $h_2>a_2'>a_2>0$.  \nopagebreak
\vspace{-2mm}
\begin{enumerate}[leftmargin=*,noitemsep]
    \item An improvement in AI's strong dimension strictly raises human wages, $w^N(a_1',a_2)>w^N(a_1,a_2)$
   \item An improvement in AI's weak dimension shrinks the set of problems that can only be solved with human input but has an ambiguous effect on the wage. It raises the wage if and only if
\[
\frac{\Pr(\boldsymbol{x}\in \Delta A)}
     {\Pr(\boldsymbol{x}\in N)}
>
\frac{\Pr(\boldsymbol{x}\in \Delta B)}
     {\Pr(\boldsymbol{x}\in H)}.
\]
It lowers the wage when the inequality is reversed and leaves the wage unchanged under equality. Either strict inequality can arise under our maintained assumptions.
   
\end{enumerate}
\end{prop}
\end{samepage}

\begin{proof} See Appendix \ref{app:prop1}. \end{proof}

The key implication is that progress in AI’s weak dimension can raise the wage even as it reduces the set of problems solvable only with human input. This possibility hinges on imperfect knowledge integration. As we show next, under perfect integration, the problems generating the favorable screening effect are themselves successful referrals, so the same improvement necessarily lowers the wage.

\subsection{Perfect Integration} \label{subsec:integration}

Under perfect integration, the wage is given by:
\[
w^I(a_1,a_2) =
\frac{\Pr(\boldsymbol{x}\in H\cup HA)}
{1-\Pr(\boldsymbol{x}\in A\cup B)} = \Pr(\boldsymbol{x}\in H\cup HA \, | \, \text{AI fails})
\]

Consider first the same improvement in AI's strong dimension studied in Section~\ref{subsec:no_integration}: $a_1$ increases to $a'_1>a_1$, while $a_2$ remains unchanged. Under no integration, this improvement raises the wage by removing the unsuccessful referrals in $\Delta A$. The same screening effect remains under perfect integration: as panel (a) of Figure~\ref{fig:4} illustrates, the problems in $\Delta A$ move from $N\setminus HA$ to $A$, so the machine no longer refers problems the organization could not previously solve.

\begin{figure}[t!]
\centering
\begin{subfigure}{.455\linewidth}
\centering
\centering
\pgfmathsetmacro{\hone}{0.35}
\pgfmathsetmacro{\htwo}{0.75}
\pgfmathsetmacro{\mone}{0.75}
\pgfmathsetmacro{\mtwo}{0.35}
\pgfmathsetmacro{\monep}{0.90}
\begin{tikzpicture}[scale=0.875]
\pgfmathsetlengthmacro\MajorTickLength{
  \pgfkeysvalueof{/pgfplots/major tick length} *1.5
}
\begin{axis}[
    ytick={\mtwo,\htwo,1},
    yticklabels={$a_2$,$h_2$,1},
    xtick={0,\hone,\mone,\monep,1},
    xticklabels={0,$h_1 \color{white}' \color{black}$,$a_1 \color{white}' \color{black}$,$a_1'$,1},
    xmin=0,xmax=1,ymin=0,ymax=1,
    axis line style=thick,
    yticklabel style={fill=white},
    ylabel={Dimension 2},
    xticklabel style={fill=white},
    xlabel={Dimension 1},
    ylabel near ticks,
    xlabel near ticks,
    ylabel style={rotate=-0},
    major tick length=\MajorTickLength,
    every tick/.style={black,thick},
    clip=false
]

\fill[pattern=north west lines,pattern color=gray!25]
    (axis cs:0,0)--(axis cs:0,\htwo)--(axis cs:\hone,\htwo)--
    (axis cs:\hone,\mtwo)--(axis cs:\mone,\mtwo)--
    (axis cs:\mone,0)--cycle;
\fill[pattern=north east lines,pattern color=ao!50]
    (axis cs:\hone,\mtwo) rectangle (axis cs:\mone,\htwo);

\fill[pattern=crosshatch,pattern color=customblue!35]
    (axis cs:\mone,0) rectangle (axis cs:\monep,\mtwo);
\fill[pattern=crosshatch,pattern color=customblue!35]
    (axis cs:\mone,\mtwo) rectangle (axis cs:\monep,\htwo);
    
\addplot[name path=ceiling,domain={0:1},opacity=0] {1};
\addplot[name path=floor,domain={0:1},opacity=0] {0};

\draw[thick,black,dotted]
    (axis cs:0,\mtwo)--(axis cs:\mone,\mtwo)--(axis cs:\mone,0);
\draw[thick,black,dotted]
    (axis cs:0,\htwo)--(axis cs:\hone,\htwo)--(axis cs:\hone,0);

\draw[thick,black] (axis cs:0,\htwo)--(axis cs:\mone,\htwo);
\draw[thick,black] (axis cs:\mone,\htwo)--(axis cs:\mone,0);

\node[font=\Large] at (axis cs:0.15,0.6) {$H$};
\node[font=\Large] at (axis cs:0.15,0.15) {$B$};
\node[font=\Large] at (axis cs:0.55,0.15) {$A$};
\node[font=\Large] at (axis cs:0.80,0.85) {$N\setminus HA$};
\node[font=\Large] at (axis cs:0.55,0.6) {$HA$};

\draw[thick,black,dashed]
    (axis cs:\mone,\mtwo)--(axis cs:\monep,\mtwo)--(axis cs:\monep,0);
\draw[thick,black,dashed]
    (axis cs:\mone,\htwo)--(axis cs:\monep,\htwo)--(axis cs:\monep,0);
\draw[thick,black,dashed,->]
    (axis cs:\mone+0.02,\mtwo+0.045)--(axis cs:\monep-0.02,\mtwo+0.045);

\node[font=\large,fill=white,inner sep=1.5pt]
    at (axis cs:{(\mone+\monep)/2},{\mtwo/2}) {\small $\Delta A$};
\node[font=\large,fill=white,inner sep=1.5pt]
    at (axis cs:{(\mone+\monep)/2},{(\mtwo+\htwo)/2}) {\small $\Delta HA$};
\end{axis}
\end{tikzpicture}
\caption*{$\quad \ \ $ (a) Improvement in AI’s Strong Dimension}

\end{subfigure}
\begin{subfigure}{.5\linewidth}
\centering
\pgfmathsetmacro{\hone}{0.35}
\pgfmathsetmacro{\htwo}{0.75}
\pgfmathsetmacro{\mone}{0.75}
\pgfmathsetmacro{\mtwo}{0.35}
\pgfmathsetmacro{\mtwop}{0.52}
\begin{tikzpicture}[scale=0.875]
\pgfmathsetlengthmacro\MajorTickLength{
  \pgfkeysvalueof{/pgfplots/major tick length} *1.5
}
\begin{axis}[
    ytick={\mtwo,\mtwop,\htwo,1},
    yticklabels={$a_2$,$a_2'$,$h_2$,1},
    xtick={0,\hone,\mone,1},
    xticklabels={0,$h_1 \color{white}' \color{black}$,$a_1 \color{white}' \color{black}$,1},
    xmin=0,xmax=1,ymin=0,ymax=1,
    axis line style=thick,
    yticklabel style={fill=white},
    ylabel={Dimension 2},
    xticklabel style={fill=white},
    xlabel={Dimension 1},
    ylabel near ticks,
    xlabel near ticks,
    ylabel style={rotate=-0},
    major tick length=\MajorTickLength,
    every tick/.style={black,thick},
    clip=false
]

\fill[pattern=north west lines,pattern color=gray!25]
    (axis cs:0,0)--(axis cs:0,\htwo)--(axis cs:\hone,\htwo)--
    (axis cs:\hone,\mtwo)--(axis cs:\mone,\mtwo)--
    (axis cs:\mone,0)--cycle;
\fill[pattern=north east lines,pattern color=ao!50]
    (axis cs:\hone,\mtwop) rectangle (axis cs:\mone,\htwo);

\fill[pattern=dots,pattern color=customred!80]
    (axis cs:0,\mtwo) rectangle (axis cs:\hone,\mtwop);
\fill[pattern=dots,pattern color=customred!80]
    (axis cs:\hone,\mtwo) rectangle (axis cs:\mone,\mtwop);
    
\addplot[name path=ceiling,domain={0:1},opacity=0] {1};
\addplot[name path=floor,domain={0:1},opacity=0] {0};

\draw[thick,black,dotted]
    (axis cs:0,\mtwo)--(axis cs:\mone,\mtwo)--(axis cs:\mone,0);
\draw[thick,black,dotted]
    (axis cs:0,\htwo)--(axis cs:\hone,\htwo)--(axis cs:\hone,0);

\draw[thick,black] (axis cs:0,\htwo)--(axis cs:\mone,\htwo);
\draw[thick,black] (axis cs:\mone,\htwo)--(axis cs:\mone,0);

\node[font=\Large] at (axis cs:0.15,0.62) {$H$};
\node[font=\Large] at (axis cs:0.15,0.15) {$B$};
\node[font=\Large] at (axis cs:0.55,0.15) {$A$};
\node[font=\Large] at (axis cs:0.80,0.85) {$N \setminus HA$};
\node[font=\Large] at (axis cs:0.55,0.62) {$HA$};

\draw[thick,black,dashed]
    (axis cs:0,\mtwop)--(axis cs:\mone,\mtwop)--(axis cs:\mone,0);
\draw[thick,black,dashed,->]
    (axis cs:\mone+0.045,\mtwo)--(axis cs:\mone+0.045,\mtwop);

\node[font=\large,fill=white,inner sep=1.5pt]
    at (axis cs:{\hone/2},{(\mtwo+\mtwop)/2}) {$\Delta B$};
\node[font=\large,fill=white,inner sep=1.5pt]
    at (axis cs:{(\hone+\mone)/2},{(\mtwo+\mtwop)/2}) {$\Delta A$};
\end{axis}
\end{tikzpicture}
\caption*{$\quad \ \ $ (b) Improvement in AI’s Weak Dimension}
\end{subfigure}
\captionsetup{justification=centering}
\caption{The Wage Effects of AI Improvements under Perfect Integration \\ \justifying 
\footnotesize \noindent \textit{Notes}. The figure shows how improvements in AI capabilities change the problem-space partition under perfect integration. In panel~(a), an increase in $a_1$ to $a_1'>a_1$ moves the problems in $\Delta A$, defined by $a_1<x_1\leq a_1'$ and $x_2\leq a_2$, from $N\setminus HA$ to $A$, and the problems in $\Delta HA$, defined by $a_1<x_1\leq a_1'$ and $a_2<x_2\leq h_2$, from $N\setminus HA$ to $HA$. In panel (b), an increase in $a_2$ to $a_2'>a_2$ moves the problems in $\Delta B$, defined by $x_1\leq h_1$ and $a_2<x_2\leq a_2'$, from $H$ to $B$, and the problems in $\Delta A$, defined by $h_1<x_1\leq a_1$ and $a_2<x_2\leq a_2'$, from $HA$ to $A$. All other regions remain unchanged.}
\label{fig:4}
\end{figure}

The improvement also generates a second positive effect. The problems in $\Delta HA$, defined by $a_1<x_1\leq a'_1$ and $a_2<x_2\leq h_2$, move from $N\setminus HA$ to $HA$. The machine continues to refer these problems, but the organization can now solve them by integrating the machine's knowledge in dimension 1 with the human's knowledge in dimension 2. The improvement therefore removes some unsuccessful referrals and turns others into successful ones. Both effects raise the wage. Thus, progress in AI's strong dimension remains labor-complementing under perfect integration.

The key difference from no integration arises when AI improves in its weak dimension: $a_2$ increases to $a'_2\in(a_2,h_2)$, while $a_1$ remains unchanged. Under no integration, the wage effect is ambiguous because the machine removes successful referrals in $\Delta B$ and unsuccessful referrals in $\Delta A$.

Under perfect integration, the transition involving $\Delta B$ is unchanged, but the problems in $\Delta A$ are successful rather than unsuccessful referrals. As panel (b) of Figure~\ref{fig:4} illustrates, before the improvement these problems lie in $HA$ and are solved by integrating human and machine knowledge; afterward, they move to $A$ and the machine solves them independently. Thus, every problem the machine stops referring is one on which human input was productive. The wage therefore falls unambiguously, so progress in AI's weak dimension is labor-substituting.

The following proposition formalizes these results:

\begin{prop}[AI Improvements - Perfect Integration] \label{prop:integration}
Suppose $a_1'>a_1>h_1$ and $h_2>a_2'>a_2>0$. \vspace{-2mm}
\begin{enumerate}[leftmargin=*,noitemsep]
    \item An improvement in AI's strong dimension strictly raises the wage, $w^I(a_1',a_2)>w^I(a_1,a_2)$.
    \item An improvement in AI's weak dimension strictly lowers the wage, $w^I(a_1,a_2')<w^I(a_1,a_2)$.
\end{enumerate}
\end{prop}

\begin{proof} See Appendix \ref{app:prop2}. \end{proof}

The polar cases of no and perfect integration studied in Propositions \ref{prop:no-integration} and \ref{prop:integration} extend naturally to imperfect integration. A simple way to model imperfect integration is to assume that the organization successfully integrates human and machine knowledge on an exogenous fraction of the problems in $HA$. It then succeeds on a given problem with probability:
\[
\Phi(\theta)
\equiv
\Pr(x\in A\cup B\cup H)
+
\theta\Pr(x\in HA),
\]
where $\theta\in[0,1]$ is the probability that a problem can be solved by the organization conditional on being in $HA$.  The corresponding wage is:
\[
w^\theta(a_1,a_2) = \frac{\Pr(x\in H)+\theta\Pr(x\in HA)} {1-\Pr(x\in A\cup B)} = (1-\theta)w^N(a_1,a_2)+\theta w^I(a_1,a_2) 
\]
Thus, holding $\theta$ fixed, the wage effect of any AI improvement is the corresponding convex combination of its effects under no and perfect integration.

\section{A General Problem-Level Decomposition} \label{sec:decomposition_measurement}

The previous sections identified screening and knowledge integration as two sources of organizational value and showed how improvements in AI capabilities affect wages through them. This section traces those wage effects to problem-level transitions among three states: solvable by AI alone, solvable only with human input, or unsolvable by the human--AI organization. This decomposition isolates a previously overlooked labor-complementing force that falls outside the standard automation--augmentation distinction.

\vspace{3mm} 

\noindent\textit{A Problem-Level Decomposition.}--- Let $\Omega$ denote the problem space and partition it into three sets: \begin{itemize}[noitemsep,leftmargin=2.25em]
    \vspace{-2mm}
    \item[$\mathcal{A}$:] Problems that AI can solve autonomously.
    \item[$\mathcal{H}$:] Problems that AI cannot solve autonomously, but the organization can solve using human input.
    \item[$\mathcal{U}$:] Problems that neither AI autonomously nor the organization with human input can solve.
\end{itemize}

In the two-dimensional economy of Sections \ref{sec:hierarchies} and \ref{sec:turing_valley}:
\[
\mathcal{A}=A\cup B,
\quad \text{and} \quad
(\mathcal{H},\mathcal{U})=
\begin{cases}
(H,N), & \text{without integration},
\\
(H\cup HA,N\setminus HA), & \text{under perfect integration}.
\end{cases}
\]

Assuming that AI improvements do not shrink AI's autonomous or organizational capabilities (i.e., they do not remove any problems from $\mathcal{A}$ or $\mathcal{A} \cup \mathcal{H}$), an improvement can move problems across the partition in three ways. First, problems move from $\mathcal{H}$ to $\mathcal{A}$ when those previously solvable only with human input become autonomously solvable by AI. Second, problems move from $\mathcal{U}$ to $\mathcal{H}$ when previously unsolvable problems become solvable with human input but not autonomously by AI. Third, problems move from $\mathcal{U}$ to $\mathcal{A}$ when previously unsolvable problems become autonomously solvable by AI. Figure \ref{fig:three_transitions} in the Introduction illustrates these three movements.

The comparative statics of Section~\ref{sec:turing_valley} admit a direct interpretation in these terms. Consider, for instance, an improvement in AI's weak dimension. Without integration, problems in $\Delta B$, which the human previously solved after referral, become autonomously solvable by AI and move from $\mathcal{H}$ to $\mathcal{A}$. This $\mathcal{H}$-to-$\mathcal{A}$ transition lowers the wage. Problems in $\Delta A$, which neither agent could previously solve, also become autonomously solvable by AI and move from $\mathcal{U}$ to $\mathcal{A}$. This $\mathcal{U}$-to-$\mathcal{A}$ transition raises the wage.

\vspace{3mm} 

\noindent\textit{Two Familiar Transitions: Automation and Augmentation.}--- To understand the wage effect of the different transitions, begin with $\mathcal{H}$-to-$\mathcal{A}$ and $\mathcal{U}$-to-$\mathcal{H}$. A movement from $\mathcal{H}$ to $\mathcal{A}$ is automation: AI autonomously takes over problems on which human input was previously necessary. The organization does not become able to solve any additional problems; instead, it becomes able to produce the same output with less human contribution. This transition substitutes for labor and lowers the wage.

A movement from $\mathcal{U}$ to $\mathcal{H}$ is augmentation: the human--AI organization becomes able to solve problems that were previously beyond its capabilities, while human input remains indispensable. This transition increases the value of the human--AI organization. Because compute is abundant, human time is the bottleneck, so the human captures this additional value through a higher wage.

\vspace{3mm} 

\noindent\textit{A New Source of Labor Complementarity.}--- A central contribution of this paper is to identify a new source of labor complementarity that becomes evident from the $\mathcal{U}$-to-$\mathcal{A}$ transition. This movement falls outside the standard automation--augmentation distinction: AI becomes able to solve problems that neither it nor the human--AI organization could solve before the improvement. It is not automation, because human input was never productive on these problems. Nor is it augmentation, because the set $\mathcal{H}$ of problems solvable with human input does not expand. The puzzle is why such progress nevertheless raises labor income.

To understand this puzzle, return to the baseline model of Sections \ref{sec:hierarchies} and \ref{sec:turing_valley} where machines tackle problems first. They refer problems in both $\mathcal{H}$ and $\mathcal{U}$ to humans. Before the improvement, problems that move from $\mathcal{U}$ to $\mathcal{A}$ consume scarce human time even though the organization ultimately fails. After the improvement, AI solves these problems before referral. The set of problems on which human input is productive remains unchanged, but the referral pool becomes more concentrated on that set. Put differently, this type of AI progress raises the marginal product of labor by allowing humans to specialize more in what they already do well.

The preceding argument may appear to depend on machines tackling problems first. As we formally show in Appendix~\ref{app:machines_second}, it does not. When machines tackle problems second, the same complementary force emerges in a slightly different form. The key is that the hierarchy changes the problem distribution faced by the machine, since it handles only problems the human cannot solve. Problems that move from $\mathcal{U}$ to $\mathcal{A}$ are therefore overrepresented in the referral stage relative to an unconditional problem draw. The AI improvement therefore raises the machine's contribution to organizational output by more than it raises its autarky return. Because compute is abundant, the additional organizational gain accrues to labor through a higher wage.

Both configurations---machines first or second---thus reveal a common principle: sequential production selects the problem pool faced by the agents tackling problems second, and a $\mathcal{U}$-to-$\mathcal{A}$ transition raises the value of operating on that selected pool. When machines move first, such a transition removes from the human's referral pool problems on which the human would fail, increasing the success rate among the problems that reach her. When humans move first, human failure selects a problem pool in which newly solvable problems are overrepresented, so a $\mathcal{U}$-to-$\mathcal{A}$ transition raises the machine's contribution inside the organization more than outside of it. 

This positive screening complementarity is orthogonal to the traditional automation and augmentation effects, and also to scale effects through which higher productivity expands output and raises labor demand. It arises within the organization---holding the output price fixed and without expanding either human capabilities or the set $\mathcal{H}$---through the screening generated by sequential production when knowledge is tacit and therefore problem types cannot be identified ex ante.

\section{Implications for Measurement}\label{sec:measurement}

The preceding analysis shows that, under sequential problem solving, the labor-income effects of AI progress depend not only on which problems require human input, but also on which problems reach humans.  This distinction has a direct implication for measurement. A growing evaluation literature compares the average performance of a human--AI system with that of AI alone \citep[e.g.,][]{cao2024manmachine,vaccaro2024combinations,merali2025scaling}. We show that this performance gap coincides with the marginal product of labor under simultaneous problem solving but can diverge under sequential production.

Let $Y^{AI}$ and $Y^{H+AI}$ denote the average success rates of AI alone and of the human--AI organization, respectively. AI alone succeeds on $\mathcal{A}$, while the organization succeeds on $\mathcal{A}\cup \mathcal{H}$. Hence,
\[ Y^{AI}=\Pr(x\in \mathcal{A}) \quad \text{and} \quad Y^{H+AI}=\Pr(x\in \mathcal{A}\cup  \mathcal{H}) \]
so the human--AI versus AI-only performance gap is:
\[ Y^{H+AI}-Y^{AI}=\Pr(x\in  \mathcal{H}) \]
The performance gap therefore equals the probability that a randomly drawn problem can be solved by the organization only with human input. Equivalently, it measures the share of the unconditional problem distribution on which human input is necessary for success.

As we formally show in Appendix~\ref{app:importance}, the performance gap coincides with the marginal product of labor when humans and machines are constrained to attempt the same problems simultaneously:
\[
w^{\mathrm{sim}}
=
\mathrm{MPL}^{\mathrm{sim}}
=
Y^{H+AI}-Y^{AI}.
\]Thus, under simultaneous problem solving, our framework reproduces the familiar equality between the human--AI performance gap and the marginal product of labor.

In our environment, however, tacit knowledge makes sequential problem solving optimal. Under this organization, one agent attempts each problem first and refers only its failures, so the second agent works on a selected pool of problems. The performance gap therefore need not coincide with the marginal product of labor.

This is easiest to see in our baseline hierarchy, in which machines occupy the bottom layer. AI attempts each problem first and refers the problems in $ \mathcal{H}\cup \mathcal{U}$ to humans. Human input produces a solution on the problems in $ \mathcal{H}$, but not on those in $\mathcal{U}$. Because AI refers a problem with probability:
\[
1-Y^{AI}
=
\Pr(x\in  \mathcal{H})+\Pr(x\in \mathcal{U}),
\]
the marginal product of labor---and thus the wage---when compute is abundant is then:
\[ w^{\mathrm{seq}} = \mathrm{MPL} = \Pr(x\in  \mathcal{H}\mid \text{AI fails}) = \frac{\Pr(x\in  \mathcal{H})}{\Pr(x\in  \mathcal{H})+\Pr(x\in \mathcal{U})} = \frac{Y^{H+AI}-Y^{AI}}{1-Y^{AI}} \]
The performance gap is therefore the numerator of the marginal product of labor, while the AI failure rate supplies the denominator. Average AI-only and human--AI performance are sufficient to recover the marginal product in the baseline hierarchy, but the performance gap alone is not. Because the marginal product is a ratio, changes in the gap need not track changes in the marginal product: a decline in the gap can be more than offset by a decline in the AI failure rate.

In Section \ref{subsec:no_integration} we saw a concrete example of this possibility. The weak-dimension improvement studied there reduces the mass of problems in $\mathcal{H}$, and therefore shrinks the performance gap. Yet it can reduce the mass of unsuccessful referrals in $\mathcal{U}$ by enough to raise the share of referrals on which human input produces a solution. The marginal product of labor can therefore rise even as the performance gap falls: human input is necessary on fewer problems overall but more productive conditional on referral. This is the measurement implication of the positive screening complementarity identified in this paper.

\section{Conclusion} \label{sec:final}

AI can acquire tacit knowledge, making it an important new player in knowledge hierarchies \citep{AIKE}. In this paper, we show that when knowledge is multidimensional, such hierarchies can create value via screening---by reducing the resources spent on problems that ultimately go unsolved---and integration---by allowing the hierarchy to solve problems that none of its members can solve independently. Our main result is that, when integration is imperfect, improving AI in dimensions where humans initially have an advantage can sometimes increase the marginal product of labor by raising the screening value of human--AI hierarchies.

To illustrate this insight as clearly as possible, we have held human capital fixed and asked how improvements in AI capabilities affect the marginal product of labor. A related question, central to recent work on upskilling and reskilling, is how those improvements change the incentives to acquire additional human capital \citep{tamayo2023reskilling}. Although a full analysis lies beyond the scope of this paper, our framework yields two observations in this regard.

First, the extent to which organizations can integrate human and AI knowledge affects which human-capital investments are valuable. Under perfect integration, additional human knowledge in a dimension where AI is already stronger is redundant unless it exceeds AI’s capability. Under imperfect integration, by contrast, humans must rely on their own knowledge in every dimension to solve some of the referrals, so investments in dimensions where AI already dominates can still be valuable.

Second, when AI tackles problems first, improvements in dimensions where humans initially have an advantage can reduce the value of existing human capital while increasing the return to further investment. By making AI autonomously solve problems on which workers’ existing knowledge was productive, such improvements lower current wages. At the same time, they concentrate the remaining referral pool on problems that additional human knowledge would make solvable, raising the wage return to acquiring that knowledge.

\newpage

\begin{appendix}
\numberwithin{equation}{section}

\begin{center} \Large \textsc{Appendix} \end{center}

\section{Simultaneous Problem Solving} \label{app:importance}

The main text focuses on sequential problem solving, under which one agent attempts each problem first and refers only its failures to the other. This appendix considers the alternative organization in which one human and one machine are constrained to attempt the same problem from the outset.\footnote{We thank Pascual Restrepo for suggesting this comparison.}

We establish four results. First, without knowledge integration, simultaneous problem solving is dominated by independent production. Second, the marginal product of labor equals the human--AI versus AI-only performance gap introduced in Section \ref{sec:measurement}. Third, conditional on simultaneous problem solving creating value, improvements in AI's strong dimension are labor-complementing, whereas improvements in its weak dimension are labor-substituting. Finally, simultaneous problem solving is always strictly dominated by a hierarchy in which machines attempt problems first.

\vspace{3mm} 

\noindent \textit{Setup}.--- Recall the partition $A$, $B$, $H$, and $N$ of the problem space introduced in Section \ref{sec:hierarchies} and the imperfect-integration technology introduced in Section \ref{subsec:integration}, where the organization successfully integrates human and machine knowledge on an exogenous fraction of the problems in $HA$. We apply the same integration technology when humans and machines attempt problems simultaneously. The organization therefore succeeds with probability:
\[
\Phi(\theta)
\equiv
\Pr(x\in A\cup B\cup H)
+
\theta\Pr(x\in HA).
\]
where $\theta\in[0,1]$ is the probability that a problem can be solved by the organization conditional on being in $HA$. The machine succeeds autonomously with probability $\Pr(x\in A\cup B)$, while the human succeeds autonomously with probability $\Pr(x\in B\cup H)$. We focus on the interior case in which:
\[
\Pr(x\in B)>0
\qquad\text{and}\qquad
0<\Pr(x\in A\cup B)<1.
\]

\vspace{1mm} 

\noindent\textit{Organizational Value under Simultaneous Problem Solving}.---Under simultaneous problem solving, one human and one machine attempt the same problem. Because compute is abundant, competition pins the machine's rental rate at its autonomous return, $r=\Pr(x\in A\cup B)$. The organization can therefore offer the human:
\[
w^{\mathrm{sim}}(\theta)
=
\Phi(\theta)-\Pr(x\in A\cup B)
=
\Pr(x\in H)+\theta\Pr(x\in HA).
\]
Simultaneous problem solving creates value relative to independent production if and only if this wage exceeds the human's autonomous return:
\begin{equation}\label{eq:theta}
w^{\mathrm{sim}}(\theta)
>
\Pr(x\in B\cup H)
\quad\Longleftrightarrow\quad
\theta\Pr(x\in HA)
>
\Pr(x\in B).
\end{equation}
Because simultaneous problem solving generates no screening benefit, integration is its only potential source of organizational value. The term $\theta\Pr(x\in HA)$ is the expected gain from solving problems that neither agent can solve independently, whereas $\Pr(x\in B)$ is the redundancy cost of using both agents on problems either could solve alone. Condition~\eqref{eq:theta} therefore requires the degree of knowledge integration to be sufficiently high for the former gain to exceed the latter cost.

For the remainder of the appendix, assume that condition~\eqref{eq:theta} holds, so that simultaneous problem solving creates value.

\vspace{3mm} 

\noindent \textit{The Marginal Product of Labor and the Performance Gap}.--- Because the organizational technology is replicable at constant returns, a marginal human forms an additional simultaneous organization with a machine that would otherwise operate autonomously. The resulting increase in expected output is:
\begin{equation} \label{eq:sim1} \operatorname{MPL}^{\mathrm{sim}} =\Phi(\theta) -\Pr(x\in A\cup B) = \Pr(x\in H)+\theta\Pr(x\in HA) = w^{\mathrm{sim}}(\theta) \end{equation}

The probability difference compares the expected success rate of the human–AI organization with that of AI alone. It is therefore exactly the human–AI versus AI-only performance gap introduced in Section \ref{sec:measurement}. Thus, under simultaneous problem solving, the performance gap equals the marginal product of labor for any degree of knowledge integration (as long as condition~\eqref{eq:theta} holds).

\vspace{3mm} 

\noindent \textit{The Wage Effects of AI Improvements}.--- Expression (\ref{eq:sim1}) also confirms the natural intuition discussed in Section \ref{sec:turing_valley}. Holding $\theta$ fixed, an improvement in AI's strong dimension expands $HA$, increasing the mass of problems on which human input is productive through knowledge integration and thus raising $w^{\mathrm{sim}}(\theta)$. By contrast, an improvement in AI's weak dimension contracts both $H$ and $HA$, reducing the mass of problems on which human input is productive and thus lowering $w^{\mathrm{sim}}(\theta)$. 

Hence, under simultaneous problem solving, progress that further differentiates AI from human knowledge is labor-complementing, whereas progress that makes AI less jagged is labor-substituting.

\vspace{3mm} 

\noindent \textit{Dominance by the Machine-Bottom Hierarchy}.---Finally, we relax the simultaneous-problem-solving constraint and compare the simultaneous organization with the machine-bottom hierarchy under the same degree of knowledge integration $\theta$ introduced in Section \ref{subsec:integration}.

Relative to independent production, the simultaneous organization creates value $\Pr(x\in H)+\theta\Pr(x\in HA)-\Pr(x\in B\cup H)$ whereas the machine-bottom hierarchy creates value $n(\boldsymbol{a})[\Pr(x\in H)+\theta\Pr(x\in HA)]-\Pr(x\in B\cup H)$. Their difference is therefore:
\[
\big[n(\boldsymbol{a})-1\big]\big[\Pr(x\in H)+\theta\Pr(x\in HA)\big]>0,
\]
where the inequality follows because $n(\boldsymbol{a})=1/[1-\Pr(x\in A\cup B)]>1$. The machine-bottom hierarchy therefore creates strictly more organizational value. Intuitively, the two organizations have the same problem-solving capabilities, but the hierarchy uses human input only after the machine fails. It thereby avoids redundant human effort on problems the machine can solve autonomously and allows each human to support multiple machines.

Simultaneous problem solving is consequently never optimal, regardless of the degree of knowledge integration. When condition (\ref{eq:theta}) fails, it is weakly dominated by independent production. When condition~(\ref{eq:theta}) holds, it creates value but is strictly dominated by sequential problem solving with machines on the bottom layer. \qed

\section{Proof of Proposition \ref{prop:no-integration}} \label{app:prop1}

Throughout, all regions refer to the pre-improvement partition. As shown in Section \ref{sec:screening}, the no-integration wage is: \[
w^N(a_1,a_2)
=
\frac{\Pr(\boldsymbol{x}\in H)}
     {1-\Pr(\boldsymbol{x}\in A\cup B)}
=
\frac{\Pr(\boldsymbol{x}\in H)}
     {\Pr(\boldsymbol{x}\in H)+\Pr(\boldsymbol{x}\in N)}.
\]

\vspace{3mm}

\noindent\textit{Improvement in AI's strong dimension.}--- The problems in $\Delta A$ move from $N$ to $A$. The mass of successful referrals is unchanged, while the referral probability falls by $\Pr(\boldsymbol{x}\in\Delta A)$. Hence, \begin{align*} w^N(a_1',a_2) &=
\frac{\Pr(\boldsymbol{x}\in H)}
     {1-\Pr(\boldsymbol{x}\in A\cup B)
      -\Pr(\boldsymbol{x}\in\Delta A)} =\frac{w^N(a_1,a_2)[1-\Pr(\boldsymbol{x}\in A\cup B)]}      {1-\Pr(\boldsymbol{x}\in A\cup B)
      -\Pr(\boldsymbol{x}\in\Delta A)}.
\end{align*}
It follows that:
\[
w^N(a_1',a_2)-w^N(a_1,a_2)
=
\frac{
w^N(a_1,a_2)\Pr(\boldsymbol{x}\in\Delta A)
}{
1-\Pr(\boldsymbol{x}\in A\cup B)
-\Pr(\boldsymbol{x}\in\Delta A)
}
>0.
\]
The inequality is strict because full support implies $\Pr(\boldsymbol{x}\in\Delta A)>0$, while the denominator is the post-improvement referral probability.

\vspace{3mm}
\noindent\textit{Improvement in AI's weak dimension.}--- The problems in $\Delta A$ move from $N$ to $A$, while those in $\Delta B$ move from $H$ to $B$. Successful referrals therefore fall by $\Pr(\boldsymbol{x}\in\Delta B)$, and total referrals fall by $\Pr(\boldsymbol{x}\in\Delta A\cup\Delta B)$. Thus, \[
w^N(a_1,a_2')
=
\frac{
\Pr(\boldsymbol{x}\in H)
-\Pr(\boldsymbol{x}\in\Delta B)
}{
\Pr(\boldsymbol{x}\in H)
+\Pr(\boldsymbol{x}\in N)
-\Pr(\boldsymbol{x}\in\Delta A)
-\Pr(\boldsymbol{x}\in\Delta B)
}.
\]
Since the pre- and post-improvement referral probabilities are strictly positive,
\begin{align*}
&w^N(a_1,a_2')>w^N(a_1,a_2)
\\
&\Longleftrightarrow\quad
\frac{
\Pr(\boldsymbol{x}\in H)-\Pr(\boldsymbol{x}\in\Delta B)
}{
\Pr(\boldsymbol{x}\in H)+\Pr(\boldsymbol{x}\in N)
-\Pr(\boldsymbol{x}\in\Delta A)
-\Pr(\boldsymbol{x}\in\Delta B)
}
>
\frac{
\Pr(\boldsymbol{x}\in H)
}{
\Pr(\boldsymbol{x}\in H)+\Pr(\boldsymbol{x}\in N)
}
\\
&\Longleftrightarrow\quad
\Pr(\boldsymbol{x}\in H)\Pr(\boldsymbol{x}\in\Delta A)
>
\Pr(\boldsymbol{x}\in N)\Pr(\boldsymbol{x}\in\Delta B)
\\
&\Longleftrightarrow\quad
\frac{\Pr(\boldsymbol{x}\in\Delta A)}
     {\Pr(\boldsymbol{x}\in N)}
>
\frac{\Pr(\boldsymbol{x}\in\Delta B)}
     {\Pr(\boldsymbol{x}\in H)}.
\end{align*}
The wage is unchanged under equality and falls when the inequality is reversed.

It remains to show that either strict inequality can arise under the maintained assumptions. Since $F$ is unrestricted apart from full support, it suffices to construct two probability assignments. In both cases, let:
\[
\Pr(\boldsymbol{x}\in B)=\frac{1}{10},\qquad
\Pr(\boldsymbol{x}\in A)=\frac{3}{10},\qquad
\Pr(\boldsymbol{x}\in H)=\frac{1}{5},\qquad
\Pr(\boldsymbol{x}\in N)=\frac{2}{5}.
\] The hierarchy is initially strictly optimal because:
\[
w^N(a_1,a_2)
=
\frac{1/5}{1/5+2/5}
=
\frac{1}{3}
>
\frac{3}{10}
=
F(\boldsymbol{h}).
\]

Fix $0<\varepsilon<1/26$. For the wage-increase construction, assign the shares:
\[
\frac{\Pr(\boldsymbol{x}\in\Delta A)}
     {\Pr(\boldsymbol{x}\in N)}
=
\frac{1}{2}+\varepsilon,
\qquad
\frac{\Pr(\boldsymbol{x}\in\Delta B)}
     {\Pr(\boldsymbol{x}\in H)}
=
\frac{1}{2}-\varepsilon.
\]
The condition in the proposition is then satisfied. Moreover,
\[
w^N(a_1,a_2')
=
\frac{1+2\varepsilon}{3-2\varepsilon}
>
\frac{1}{3},
\]
so the wage rises, and the hierarchy remains strictly optimal.

For the wage-decrease construction, simply swap the two normalized shares:
\[
\frac{\Pr(\boldsymbol{x}\in\Delta A)}
     {\Pr(\boldsymbol{x}\in N)}
=
\frac{1}{2}-\varepsilon,
\qquad
\frac{\Pr(\boldsymbol{x}\in\Delta B)}
     {\Pr(\boldsymbol{x}\in H)}
=
\frac{1}{2}+\varepsilon.
\]
The inequality in the proposition is now reversed, and the post-improvement wage is
\[
w^N(a_1,a_2')
=
\frac{1-2\varepsilon}{3+2\varepsilon}.
\]
The restriction $\varepsilon<1/26$ ensures that:
\[
\frac{3}{10}
<
\frac{1-2\varepsilon}{3+2\varepsilon}
<
\frac{1}{3}.
\]
Thus, the wage falls, but the hierarchy remains strictly optimal.

Under either construction, the post-improvement referral probability exceeds $1/4$, so $\mu>4$ guarantees compute abundance both initially and after the improvement. Finally, the strict capability inequalities imply that all relevant regions have positive area. A strictly positive density that is constant on each of:
\[
B,\quad A,\quad \Delta B,\quad H\setminus\Delta B,\quad
\Delta A,\quad N\setminus\Delta A
\]
implements either probability assignment and has full support. \qed

\section{Proof of Proposition \ref{prop:integration}} \label{app:prop2}

Throughout, all regions refer to the pre-improvement partition. As shown in Section~\ref{sec:integration}, the perfect-integration wage is: \[
w^I(a_1,a_2)
=
\frac{\Pr(\boldsymbol{x}\in H\cup HA)}
     {1-\Pr(\boldsymbol{x}\in A\cup B)}.
\]

\vspace{3mm}
\noindent\textit{Improvement in AI's strong dimension.}--- The problems in $\Delta A$ leave the referral pool, while those in $\Delta HA$ remain referrals but become solvable by the organization. Therefore,
\begin{align*}
w^I(a_1',a_2)
&=
\frac{
\Pr(\boldsymbol{x}\in H\cup HA)
+\Pr(\boldsymbol{x}\in\Delta HA)
}{
1-\Pr(\boldsymbol{x}\in A\cup B)
-\Pr(\boldsymbol{x}\in\Delta A)
}
\\
&=
w^I(a_1,a_2)
\frac{1-\Pr(\boldsymbol{x}\in A\cup B)}
     {1-\Pr(\boldsymbol{x}\in A\cup B)
      -\Pr(\boldsymbol{x}\in\Delta A)}
+
\frac{
\Pr(\boldsymbol{x}\in\Delta HA)
}{
1-\Pr(\boldsymbol{x}\in A\cup B)
-\Pr(\boldsymbol{x}\in\Delta A)
}.
\end{align*}
Hence,
\[
w^I(a_1',a_2)-w^I(a_1,a_2)
=
\frac{
\Pr(\boldsymbol{x}\in\Delta HA)
+
w^I(a_1,a_2)\Pr(\boldsymbol{x}\in\Delta A)
}{
1-\Pr(\boldsymbol{x}\in A\cup B)
-\Pr(\boldsymbol{x}\in\Delta A)
}
>0.
\]
Full support implies:
\[
\Pr(\boldsymbol{x}\in\Delta A)>0
\qquad\text{and}\qquad
\Pr(\boldsymbol{x}\in\Delta HA)>0,
\]
while the denominator is the post-improvement referral probability.

\vspace{3mm}
\noindent\textit{Improvement in AI's weak dimension.}--- Under perfect integration, all problems in $\Delta A\cup\Delta B$ are successful referrals before the improvement and leave the referral pool afterward. It follows that:
\[
w^I(a_1,a_2')
=
\frac{
\Pr(\boldsymbol{x}\in H\cup HA)
-\Pr(\boldsymbol{x}\in\Delta A\cup\Delta B)
}{
1-\Pr(\boldsymbol{x}\in A\cup B)
-\Pr(\boldsymbol{x}\in\Delta A\cup\Delta B)
}.
\]
Using:
\[
\Pr(\boldsymbol{x}\in H\cup HA)
=
w^I(a_1,a_2)
\bigl[1-\Pr(\boldsymbol{x}\in A\cup B)\bigr],
\]
we obtain:
$$
w^I(a_1,a_2')-w^I(a_1,a_2)=
-\frac{
\bigl[1-w^I(a_1,a_2)\bigr]
\Pr(\boldsymbol{x}\in\Delta A\cup\Delta B)
}{
1-\Pr(\boldsymbol{x}\in A\cup B)
-\Pr(\boldsymbol{x}\in\Delta A\cup\Delta B)
}.
$$
Moreover,
\[
1-w^I(a_1,a_2)
=
\frac{
\Pr(\boldsymbol{x}\in N\setminus HA)
}{
1-\Pr(\boldsymbol{x}\in A\cup B)
}
>0.
\]
Full support also implies $\Pr(\boldsymbol{x}\in\Delta A\cup\Delta B)>0$, while the denominator is the post-improvement referral probability. Therefore,
\[
w^I(a_1,a_2')-w^I(a_1,a_2)<0.
\]\qed

\section{Alternative Ordering: Machines Tackle Problems Second} \label{app:machines_second}

The baseline model assumes that machines tackle problems first and refer their failures to humans. This appendix shows that the positive screening complementarity does not depend on that ordering. When humans tackle problems first, their failures select the pool of problems on which machines operate. An AI improvement can therefore be more valuable inside the hierarchy than in autarky, generating the same screening complementarity as in the baseline model.

To isolate this mechanism, we assume no knowledge integration and abundant compute, and we restrict attention to AI improvements for which the hierarchy in which humans tackle problems first is strictly optimal both before and after the improvement. Its wage therefore coincides with the equilibrium wage both before and after the improvement. Let $q_h\equiv 1-F(\bh)=\Pr(\boldsymbol{x}\in A\cup N)$ denote the probability that the human fails. Each human pursues one production opportunity and refers the problem only upon failure. Since each referral requires one unit of machine time, the hierarchy uses $q_h$ units of machine time per human.

The organization solves the problems in $A\cup B\cup H$, so its expected output per human is $\Phi^N(\bh,\ba)=F(\bh)+\Pr(\boldsymbol{x}\in A)$. Compute abundance pins the rental rate of machine time at the machine's autarky return $F(\ba)$. The wage offered by the hierarchy is therefore \[ w(a_1,a_2)=
\Phi^N(\bh,\ba)-q_hF(\ba) = F(\bh)+ q_h\left[ \Pr(\text{AI succeeds}\mid\text{human fails}) - F(\ba) \right]. \]

This expression isolates the hierarchy's screening value. A machine operating in autarky faces an unconditional problem draw, whereas a machine inside the hierarchy is used only after the human fails. The hierarchy creates value when this selected pool is more favorable to the machine than an unconditional draw.

Consider now an improvement in AI's weak dimension from $a_2$ to $a_2'\in(a_2,h_2)$, holding $a_1$ fixed. As in panel (b) of Figure \ref{fig:3}, the problems in $\Delta A$ move from $N$ to $A$, while those in $\Delta B$ move from $H$ to $B$. Because human capabilities do not change, the referral probability $q_h$, and hence the amount of machine time used per human, remains unchanged.

The improvement raises organizational output on the problems in $\Delta A$: the human cannot solve these problems, but the improved machine can solve them after referral. It does not raise organizational output on the problems in $\Delta B$, because the human already solves them before the machine is called upon. The machine's autarky return, however, rises on both sets. The resulting wage change is:
\[ \begin{aligned}
w(a_1,a_2')-w(a_1,a_2)
&=
\Pr(\boldsymbol{x}\in\Delta A)
-
q_h\Pr(\boldsymbol{x}\in\Delta A\cup\Delta B) \\
&=
F(\bh)\Pr(\boldsymbol{x}\in\Delta A)
-
\bigl[1-F(\bh)\bigr]
\Pr(\boldsymbol{x}\in\Delta B).
\end{aligned}
\]

Consequently, the wage rises if and only if:
\begin{equation}
\label{eq:app_top_weak_condition}
\frac{\Pr(\boldsymbol{x}\in\Delta A)}
     {1-F(\bh)}
>
\frac{\Pr(\boldsymbol{x}\in\Delta B)}
     {F(\bh)}.
\end{equation}Equivalently, the improvement affects a larger share of the problems on which the human fails than of those on which the human succeeds. Because machines inside the hierarchy are used only after human failure, every problem in $\Delta A$ reaches the machine, whereas no problem in $\Delta B$ does. The improvement therefore raises machine productivity inside the hierarchy through $\Delta A$, while it raises the machine's autarky return---and hence its rental rate---through both $\Delta A$ and $\Delta B$. Condition~\eqref{eq:app_top_weak_condition} states that the productivity gain on the selected referral pool exceeds the associated increase in rental costs. Since compute is abundant, the resulting organizational gain accrues to the human through a higher wage.

The mechanism is the mirror image of the one operating in the baseline hierarchy. When machines tackle problems first, machine failures select the problems faced by humans. When humans tackle problems first, human failures select the problems faced by machines. In either ordering, sequential production changes the problem distribution faced by the second-stage agent. Positive screening complementarity arises when an AI improvement is more valuable on this selected distribution than on an unconditional problem draw. \qed \end{appendix}

{\linespread{1.3} 
\bibliographystyle{ecta}
\bibliography{Turing}}

\end{document}